\documentclass[aps,floatfix,prl,twocolumn,superscriptaddress,groupedaddress]{revtex4}
\usepackage{graphicx,amsmath,bm,floatrow}
\usepackage[caption=false]{subfig}
\usepackage{color}

\begin{document}

\title{Tunable plasmon-enhanced birefringence in ribbon array of anisotropic 2D materials}



\author{{ Kaveh Khaliji$^{1}$, Arya Fallahi$^2$, Luis Martin-Moreno$^{3}$, and Tony Low$^{1}$} \\
{\small \em $^1$ Department of Electrical and Computer Engineering, University of Minnesota, Minneapolis, MN 55455, USA\\
$^2$ DESY-Center for Free-Electron Laser Science, Notkestrasse 85, D-22607 Hamburg, Germany \\
$^3$ Instituto de Ciencia de Materiales de Aragon and Departamento de Fisica de la Materia Condensada,\\
CSIC-Universidad de Zaragoza, E-50009 Zaragoza, Spain\\}}


\begin{abstract}
We explore the far-field scattering properties of anisotropic 2D materials in ribbon array configuration. Our study reveals the plasmon-enhanced linear birefringence in these ultrathin metasurfaces, where linearly polarized incident light can be scattered into its orthogonal polarization or be converted into circular polarized light. We found wide modulation in both amplitude and phase of the scattered light via tuning the operating frequency or material's anisotropy and develop models to explain the observed scattering behavior.
\end{abstract}


\maketitle

\emph{Introduction} -- Metal-based metasurfaces can allow ways to manipulate light not possible in natural media, such as anomalous reflection and refraction \citep{yu2011light}, photonic spin Hall effect \citep{yin2013photonic}, sub-wavelength imaging \citep{khorasaninejad2016metalenses}, among many other optical phenomena \citep{kildishev2013planar, yu2014flat, chen2016areview}. Although, one can tailor the optical response of metallic metasurfaces with geometry and choice of constituent metals, such metasurfaces cannot be cast in reconfigurable photonics, where real time control over the designated functionality is demanded \citep{zheludev2012metamaterials}. As alternate platforms, two dimensional (2D) materials \citep{low2016polaritons, basov2016polaritons}, such as graphene, which allow for active modulation of optical properties via electrical \citep{ju2011graphene}, chemical \citep{yan2013damping}, and optical \citep{ni2016ultrafast} means, garner attention as natural material choice for application in tunable planar photonics \citep{koppens2014photodetectors, sun2016optical, mak2016photonics}.

The linear birefringent effect, which denotes direction-dependent phase accumulation of linearly polarized light, relies on the anisotropic property of the host medium \citep{orfanidis2002electromagnetic}. In metasurfaces, the latter can be achieved through \emph{artificial} manipulation of the surface itself (with anisotropic doping \citep{huidobro2016graphene} or patterning \citep{hipolito2012enhanced, liu2016localized}) or its surroundings (through integration with an array of anisotropic metallic or dielectric patches \citep{li2016optical, vakil2011transformation}). Alternately, with the recent isolation of anisotropic 2D materials \citep{li2014black, liu2014phosphorene, chenet2015in, island2016titanium}, one can exploit the \emph{inherent} anisotropy of the crystal lattice to induce the phase anisotropy \citep{lan2016visualizing, qiao2014high}. In homogeneous form, such 2D materials with anisotropic \citep{wang2015highly, aslan2015linearly} and hyperbolic \citep{nemilentsau2016anisotropic, correas2016black} polaritonic properties, can be regarded as ideal material platforms to be used as ultra-thin linearly birefringent retarders. 

In this work, we study light scattering properties in anisotropic 2D materials and show how plasmon excitation in ribbon array (RA) configuration can enable a wide range of control over the amplitude, phase, and polarization state of the scattered light. Through inspection of various scenarios, we found that the mere rotation of the array plane relative to the incident field polarization, or modulation of material Drude weights through control of its carrier density or effective mass, can be adopted to drastically tune the RA optical response.

\begin{figure*}
\floatbox[{\capbeside\thisfloatsetup{capbesideposition={right,center},capbesidewidth=5.0cm}}]{figure}[1.4\FBwidth]{\caption{(a) The schematic illustration of plane wave interaction with homogeneous anisotropic 2D material. Contour plots of the reflected (b) power and (c) absolute rotation angle (in degrees) versus frequency and angular detuning. (d) and (e) are similar to (b) and (c) for the transmitted field.}\label{homo-2D}}
{\includegraphics[trim = 0mm 0mm 0mm 0mm, clip, width=1.0\linewidth]{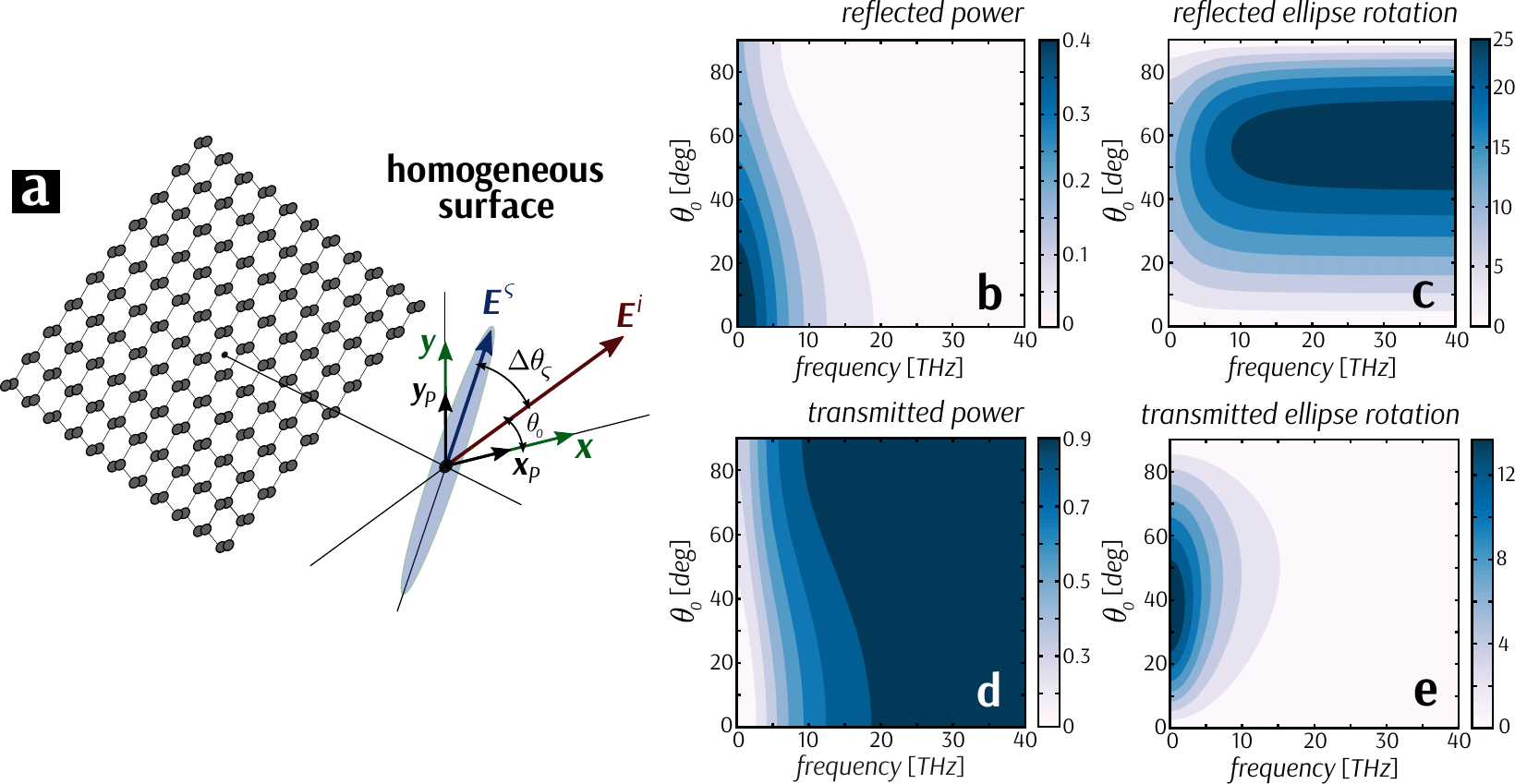}}
\end{figure*}

\emph{Homogeneous anisotropic surface} -- To avoid additional scattering effects due to index contrast, we focus our study on a free-standing anisotropic surface. The 2D crystal resides on the $x-y$ plane, where $x$ and $y$ are set to be along the lattice high-symmetry directions, i.e., $x = x_{p}$ and $y = y_{p}$, with $x_{p}$ denoting the crystal axis with the highest static conductivity; see the schematic illustration in Fig.\,\ref{homo-2D}(a). 

For a \emph{linearly}-polarized plane wave impinging normally on the anisotropic 2D lattice, the scattered fields, in general, are \emph{elliptically}-polarized plane waves with the scattered power, given as \citep{balanis2012advanced}:
\begin{equation}
\label{power-homo}
|\varsigma|^{2} = \frac{|\varsigma_{x}|^{2}+|\varsigma_{y}|^{2} \tan^{2} \theta_{0}}{1+ \tan^{2} \theta_{0}},
\end{equation}
and the ellipse major-axis rotation of:
\begin{equation}
\label{rot-angle-homo}
\begin{split}
\Delta \theta_{\varsigma} & = \frac{1}{2} \arctan \left(~\mathcal{A}^{\varsigma},\mathcal{B}^{\varsigma}~\right) - \theta_{0} \\
\mathcal{A}^{\varsigma} & = 2\left|\varsigma_{x} \varsigma_{y} \tan \theta_{0} \right| \cos \psi_{\varsigma} \\
\mathcal{B}^{\varsigma} & = \left|\varsigma_{x}\right|^{2}-\left|\varsigma_{y}\right|^{2}\tan^{2} \theta_{0}.
\end{split}
\end{equation}
In Eqs. (\ref{power-homo}) and (\ref{rot-angle-homo}), $\varsigma \rightarrow r (t)$ for the reflected (transmitted) wave and $\arctan \left( \cdot,\cdot \right)$ is the four quadrant inverse tangent function. $\theta_{0}$ is the angular detuning and denotes the angle between the incident polarization vector and $x$-axis; see Fig.\,\ref{homo-2D}(a). $\varsigma_{j}$ with $j \in \{x,y\}$ is the scattering amplitude, defined as the ratio of the scattered field along the $j$-axis over the field component of the incident wave parallel to the same axis. For the homogeneous surface, the latter can be calculated through:
\begin{equation}
\label{amplitude-homo}
r_{j} = \frac{-\sigma^{p}_{jj}}{2Y_{0}+\sigma^{p}_{jj}}, ~~~ t_{j} = \frac{2Y_{0}}{2Y_{0}+\sigma^{p}_{jj}},
\end{equation}
where $Y_{0}=\sqrt{\epsilon_{0}/\mu_{0}}$ is the intrinsic admittance of the free space. A Drude-like expression is used to model the dynamic conductivity of the anisotropic surface:
\begin{equation}
\label{cond-Drude-homo}
\sigma^{p}_{jj}(\omega)=\frac{i\mathcal{D}_{j}}{\omega+i\delta/ \hbar},
\end{equation}
where $\delta$ accounts for the finite carrier lifetime, taken to be $10\,$meV in this work, and $\mathcal{D}_{j}$ denotes the Drude weight along the corresponding axis, $j$ \citep{low2014plasmons}. We note that for homogeneous surface, off diagonal conductivity elements are zero in the basis of principal axes. The Drude model would suffice for terahertz frequencies since the band gaps of known anisotropic 2D materials are in the mid-infrared to visible \citep{tran2014layer, tongay2014monolayer, jin2015single}. Take for instance, a $10\,$nm black phosphorous (BP) film, with a doping of $0.3\,$eV, yields the pair of Drude weights $\mathcal{D}_{x}=162$ and $\mathcal{D}_{y}=59\,$GHz/$\Omega$ (see supplementary material). Throughout this study, unless mentioned otherwise, these Drude weights are assumed. The quantity $\psi_{\varsigma}=\angle \varsigma_{y} - \angle \varsigma_{x}$, measures the phase retardation between the components of the scattered field along the two coordinate axes. To determine the polarization type of the scattered field, we compute the ellipticity angle, $\phi_{\varsigma}$, defined as: $\tan 2\phi_{\varsigma} \, = \, \tan \psi_{\varsigma} \sin 2\left(\Delta \theta_{\varsigma} + \theta_{0}\right)$, where $\phi_{\varsigma} = 0^{\circ}/45^{\circ}$ identifies the scattered field as a plane wave with linear/circular field polarization \citep{balanis2012advanced}.
  
The homogeneous surface response and its dependence on frequency and angular detuning are summarized in Figs.\,\ref{homo-2D}(b)-(e). When incident light is linearly-polarized along the principal axes, the scattered field remains linear with zero rotation of polarization plane. The corresponding power in this case is simplified to: $|\varsigma|^2 = |\varsigma_{x/y}|^{2}$ for $\theta_{0}=0^{\circ}/90^{\circ}$ angular detuning. However, when the incident polarization is not aligned with the crystal high-symmetry axes, the anisotropic Drude absorption renders the scattered fields to be of elliptical form with nonzero ellipse rotation. The observed trends in this scenario suggest a trade-off between the scattered power and $\Delta \theta_{\varsigma}$, where one can identify a maximum for ellipse rotation angle, $\Delta \theta^{\textsf{max}}_{\varsigma}$ at a particular angular detuning, $\theta^{\textsf{max},\varsigma}_{0}$. These quantities can be well approximated by (see the supplementary material):
\begin{equation}
\begin{split}
\label{rot-angle-max-homo}
& ~~ \theta^{\textsf{max},\varsigma}_{0} \simeq \arctan \sqrt{\left|\frac{\varsigma_{x}}{\varsigma_{y}}\right|}\\
\Delta \theta^{\textsf{max}}_{\varsigma} \simeq & \arctan \left( \frac{1}{2} \left( \sqrt{\left|\frac{\varsigma_y}{\varsigma_x}\right|} -\sqrt{\left|\frac{\varsigma_x}{\varsigma_y}\right|}~\right) \right).
\end{split}
\end{equation}
From Figs.\,\ref{homo-2D}(c) and (e), the maximum rotation angle is higher for the reflected wave compared to that of the transmitted wave. Moreover, $\theta^{\textsf{max},r}_{0}>\theta^{\textsf{max},t}_{0}$. Equations (\ref{amplitude-homo}) and (\ref{rot-angle-max-homo}) along with $\left|\sigma^{p}_{xx}\right|>\left|\sigma^{p}_{yy}\right|$ can be invoked to justify these observations.

From Eq. (\ref{rot-angle-max-homo}), the linear birefringent effect and its induced ellipse rotation is directly dependent on the ratio of the scattered amplitudes. Thus, to engineer the ratio, one may pattern the homogeneous 2D surface into periodic array of microribbons \citep{zhao2016growth, island2014ultrahigh}. In the array geometry, the scattered amplitude, perpendicular to the ribbons, can be enhanced through excitation of localized plasmons. We next, elaborate how patterning can play a role in enhancing the linear birefringence.


\begin{figure*}
\centering
\includegraphics[width=6.5in]{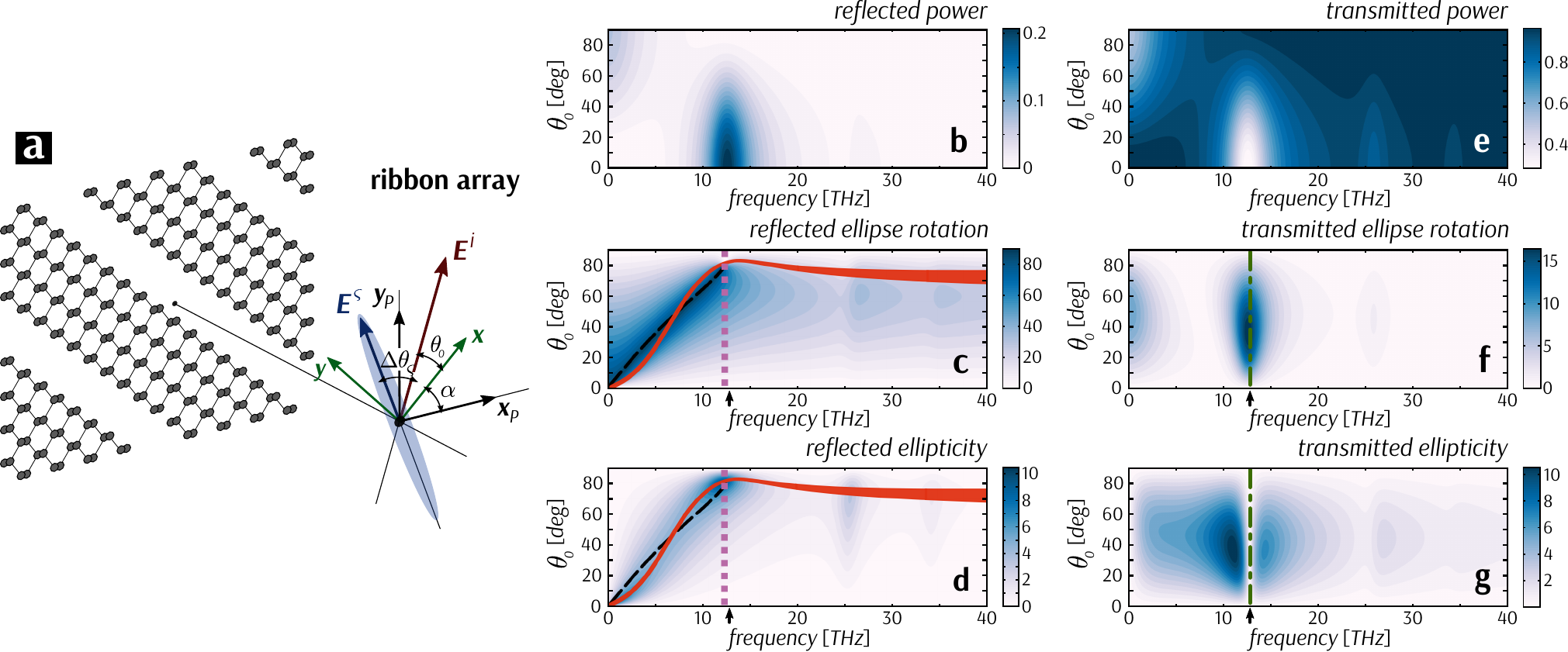}
\caption{(a) The schematic representation of ellipse major-axis rotation in the fields scattered of microribbon array. The angular detuning and frequency dependence of the (b) power, (c) absolute rotation angle (in degrees), and (d) absolute ellipticity (in degrees) of the reflected field for RA with $\alpha = 0^{\circ}$, $L = 6 \mu$m, and $f = 0.5$. (e), (f), and (g) are similar to (b), (c), and (d), respectively, for the transmitted field. In (c) and (d), the black dashed, red solid, and purple dotted lines denote data points corresponding to $\psi_{r} = 90^{\circ}$, $E^{r}_{x} = E^{r}_{y}$, and $|\Delta \theta_{r}| = 90^{\circ}$, respectively, obtained with the analytical method. In (f) and (g), green dot-dashed lines show the analytical results for $\psi_{t} = 0^{\circ}$. The black arrows denote the fundamental plasmon frequency obtained from the analytical model. The analytic expressions clearly reproduce the key features in full-wave numerical results.}
\label{RA}
\end{figure*}

\emph{Periodic array of microribbons} -- To properly model the plane wave interaction with patterned anisotropic metasurfaces, one needs to resort to numerical approaches for solving the Maxwell's equations in conjunction with the appropriate boundary conditions. For this purpose, we use the periodic method of moments technique, which is widely adopted for the simulation of patterned conductive metasurfaces in the past decades \citep{mittra1988techniques, munk2005}. The method offers straightforward implementation of the anisotropic conductivity and provides a versatile platform for analyzing the plane wave interaction \citep{fallahi2012design, fallahi2010optimal}. In line with the study we conduct for the homogeneous surface, here again we are interested in free standing array under normal illumination.

In conjunction to the numerical results, we also derived an approximate analytical model, which helps develop physical intuition on the diffraction problem. For metal grid reflectors at a dielectric boundary, using a transmission line analogy, an effective conductivity tensor can be defined as: ${\bf{\underline{\sigma}} ^{t}}= f ( {\bf{\underline{Z}}}_m + f {\bf{\underline{Z}}}_g )^{-1}$, where $f$ is the filling factor and ${\bf{\underline{Z}}}_m$ (${\bf{\underline{Z}}}_g$) is the impedance tensor of metal (gap) region \citep{fallahi2012manipulation}. For the metallic segment, the impedance tensor written in $x-y$ coordinate system can be calculated as: ${\bf{\underline{Z}}}^{-1}_m = \mathcal{M}_{\alpha} {\bf{\underline{\sigma}}}^{p} \mathcal{M}_{\alpha}^{{-1}}$, where $\mathcal{M_{\alpha}}$ is the 2D rotation matrix and ${\bf{\underline{\sigma}}}^{p}$ is the principal conductivity tensor with its nonzero elements given in Eq. (\ref{cond-Drude-homo}). Note that for the array configuration, $x$ and $y$ axes are chosen to be parallel and perpendicular to the strips. Thus, depending on the patterning angle, $\alpha$, these axes may not coincide with the crystal principal directions; see Fig.\,\ref{RA}(a). The coupling impedance tensor, which models the gap portion, is computed through ${\bf{\underline{Z}}}_g=\frac{i}{\omega \mathcal{C}_{c}}{\widehat{x} \widehat{x}}$, where, $\mathcal{C}_{c}=\frac{2\epsilon_{0}L}{\pi}\ln \left( \csc \left( \pi \left( 1- f \right) /2 \right) \right)$ denotes the near-field coupling capacitance and $L$ is the grid period \citep{whitbourn1985equivalent-circuit}.

With the homogenized-surface description of the RA at our disposal, the scattered amplitudes of the patterned surface are obtained as (see the supplementary material):
\begin{equation}
\label{amplitude-array}
\begin{split}
r_{x} = - & \mbox{\large\(\Sigma\)}^{-1} \big[ \sigma^{t}_{xx} ( 2Y_{0}+\sigma^{t}_{yy} ) + \sigma^{t}_{xy} ( 2Y_0 \tan \theta_0 - \sigma^{t}_{xy} ) \big],\\
r_{y} = - & \mbox{\large\(\Sigma\)}^{-1} \big[ \sigma^{t}_{yy} ( 2Y_{0}+\sigma^{t}_{xx} ) + \sigma^{t}_{xy} ( 2Y_0 \cot \theta_0 - \sigma^{t}_{xy} ) \big],\\
& t_{x} = \mbox{\large\(\Sigma\)}^{-1} \big[2Y_{0}( 2Y_{0}+\sigma^{t}_{yy}-\sigma^{t}_{xy}\tan \theta_{0}) \big],\\
& t_{y} = \mbox{\large\(\Sigma\)}^{-1}\big[2Y_{0}( 2Y_{0}+\sigma^{t}_{xx}-\sigma^{t}_{xy}\cot \theta_{0} ) \big],
\end{split}
\end{equation}
where, $\Sigma = ( 2Y_0 + \sigma^{t}_{xx} ) ( 2Y_0 + \sigma^{t}_{yy} )-( \sigma^{t}_{xy} )^2$. These quantities are substituted in Eqs. (\ref{power-homo}) and (\ref{rot-angle-homo}), to compute the scattered power and rotation angle of the RA.

For the array geometry, we first consider the simplest setup: ribbons are patterned along the crystal high-symmetry directions ($\alpha \in  \{0^{\circ}, 90^{\circ}\}$) and the incident field has polarization parallel ($\theta_{0} = 90^{\circ}$) or perpendicular ($\theta_{0} = 0^{\circ}$) to the ribbon axes. In terms of ellipticity and rotation angle, the array response resembles the homogeneous surface with $\theta_{0} \in \{0^{\circ}, 90^{\circ}\}$, as scattered fields exhibit neither polarization rotation nor linear-elliptic polarization conversion. The scattered power, however, deviates from that of the 2D surface. For parallel polarization, the reflected (transmitted) power decreases (increases) almost uniformly with patterning, in accordance to the reduced filling factor of the ribbons. For perpendicular incidence, however, the reflected (transmitted) spectra manifests a peak (dip) due to excitation of the localized plasmons. 

\begin{figure*}
\floatbox[{\capbeside\thisfloatsetup{capbesideposition={right,center},capbesidewidth=5.5cm}}]{figure}[1.3\FBwidth]{\caption{The reflected wave (a) power, (b) ellipse rotation angle (in degrees), and (c) ellipticity (in degrees) as functions of $\gamma$ and frequency for RA with ($\alpha = 0^{\circ}$, $\theta_{0} = 35^{\circ}$). (d), (e), and (f) are similar to (a), (b), and (c), respectively, for the transmitted field of the array with ($\alpha = 25^{\circ}$, $\theta_{0} = 35^{\circ}$). For all panels: $L = 6 \mu$m and $f = 0.5$.}\label{tunability}}
{\includegraphics[trim = 0mm 0mm 0mm 0mm, clip, width=0.98\linewidth]{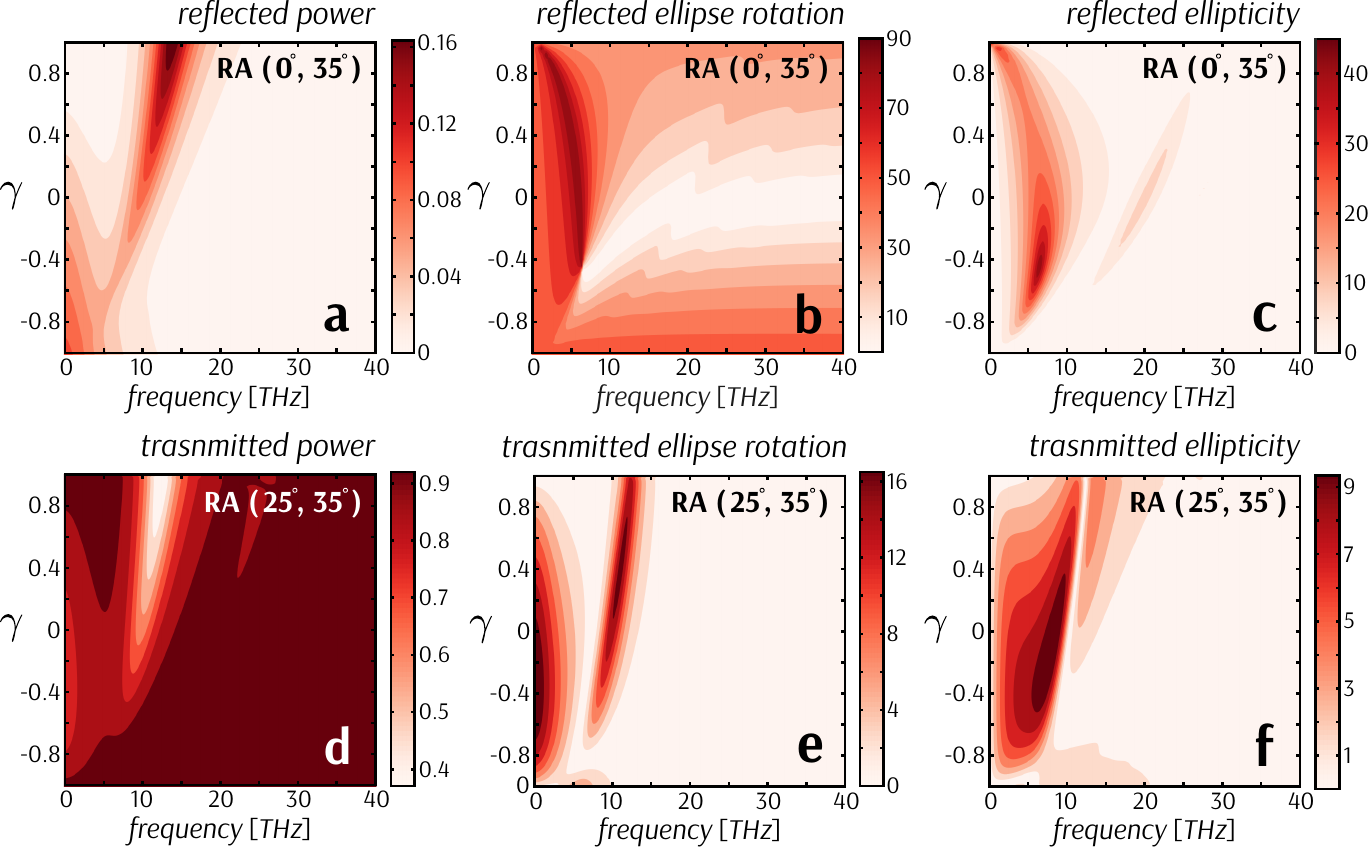}}
\end{figure*}
   
Next, we proceed to the case where $\theta_{0} \notin \{0^{\circ}, 90^{\circ}\}$, while the array is still assumed to be cut along the high-symmetry axes, i.e., $\alpha \in \{0^{\circ}, 90^{\circ}\}$. We discuss mainly the case where, $\alpha = 0^{\circ}$, for which the conductivity elements are simplified to: $\left( \sigma^{t}_{xx} \right)^{-1}_{}=\left( f\sigma^{p}_{xx} \right)^{-1}_{}-\left( i\omega \mathcal{C}_{c}\right)^{-1}_{}$, $\sigma^{t}_{xy} = \sigma^{t}_{yx} = 0$, and $\sigma^{t}_{yy} = f \sigma^{p}_{yy}$. As shown in Figs.\,\ref{RA}(b) and (e), the power spectrum for $\theta_{0} \in (0^{\circ},90^{\circ})$, still exhibit the plasmon resonance, although its strength, in terms of the reflection peak or transmission drop decreases with angular detuning. Focusing on the characteristic angles, the reflected field ellipse undergoes major axis rotation as large as 90 degree at pairs of angular detuning and frequency which satisfy the following relation:
\begin{equation}
\label{90-degree-rot-refl}
\begin{split}
\tan^{2^{}} \theta_{0} \, = \, &\frac{\cos \psi_{r} - \left|r_{x}/r_{y}\right|}{\cos \psi_{r} - \left|r_{y}/r_{x}\right|},
\end{split}
\end{equation}
where $r_{x}$ and $r_{y}$ are given in Eq. \ref{amplitude-array}. As shown in the Figs.\,\ref{RA}(c) and (d), the frequency for which $\Delta \theta_{r}=90^{\circ}$, increases with angular detuning, until it reaches a point where the reflected field acquires circular polarization. Quantitatively, the latter can be tracked with the following criteria: $\Re \left\lbrace r_{y}/r_{x} \right\rbrace =0$ and $\left|\Im \left\lbrace r_{x}/r_{y} \right\rbrace \right|=\tan \theta_{0}$, corresponding to 90 degree phase retardation and amplitude-equality of the reflected field components along the two coordinate axes. For the transmitted wave, as illustrated in Figs.\,\ref{RA}(f) and (g), the field maximum rotation occurs at plasmon resonance. Furthermore, irrespective of $\theta_{0}$, transmitted field remains linearly polarized at the plasmon frequency, for which the equivalent condition reads as: $\Im \left\lbrace t_{y}/t_{x} \right\rbrace =0$, with $t_{x}$ and $t_{y}$ as given in Eq. \ref{amplitude-array}.

As a final remark in this section, we emphasize that various configurations may be adopted to enhance the linear birefringence through coupling parallel and perpendicular responses of the RA. These include scenarios when (1) ribbons are patterned along the crystal axes, but illuminated with $\theta_{0} \notin \{0^{\circ},90^{\circ}\}$, (2) incident field is polarized along the principal axes, but the array is cut with $\alpha \notin\{0^{\circ},90^{\circ}\}$ or (3) a hybrid scheme of (1) and (2) is incorporated. Although, our discussion so far covers mostly scheme (1), the analytical framework developed here can be used for other configurations as well. As we shall discuss in the following section, these schemes provide additional degrees of freedom in tuning the operating point of metadevices based on these anisotropic surfaces.

\emph{Tunability of the linear birefringent effect} -- To assess the impact of anisotropy on the electromagnetic response, we define $\gamma=\left(\mathcal{D}_{x}-\mathcal{D}_{y}\right)/(\mathcal{D}_{x}+\mathcal{D}_{y})$ as a measure to quantify the degree of anisotropy in the material. We vary $\gamma$ by increasing $\mathcal{D}_{x}$ or $\mathcal{D}_{y}$, while leaving the sum unchanged: $\mathcal{D}_{x}+\mathcal{D}_{y}=200$ GHz/$\Omega$. $\gamma=0$ implies the isotropic material, while $|\gamma| = 1$ denotes extreme anisotropy where conductivity is zero along one of the crystal axes. We stress that the $\gamma$-factor depends primarily on the Fermi level and crystal effective masses, quantities which can be tuned in experiments with electrostatic gating \citep{li2014black}, in-situ doping \citep{kim2015observation}, or strain engineering \citep{roldan2015strain}.

According to Figs.\,\ref{tunability}(a) and (d), the plasmon resonance exhibit a blue shift as $\gamma$ approaches unity. Furthermore, a red shift in plasmon frequency can be observed with increasing the patterning angle. These trends are consistent with $\sqrt{\mathcal{D}_{x} \cos^{2}\alpha + \mathcal{D}_{y} \sin^{2}\alpha}$\, dependence, calculated using the plasmon dispersion of the anisotropic homogeneous surface in quasi-electrostatic limit \citep{low2014plasmons, liu2016localized}. Comparing the reflected mode results in Figs.\,\ref{RA}(c) and (d) with those in Figs.\,\ref{tunability}(b) and (c), it is apparent that the frequency range at which linear-circular polarization conversion takes place, can be widely tuned with modulation of the anisotropy parameter and angular detuning. Furthermore, from panels (d)-(f) of Fig.\,\ref{tunability}, the transmitted field polarization, although exhibit nonzero rotation, it remains linear at plasmon frequencies. The tunability of the plasmon resonance with $\gamma$ and patterning angle, thus, renders the RA configuration a dynamic polarization rotator in the transmitted mode.


\emph{Conclusion} -- In summary, through full-wave calculation and intuitive analytical formulation, we discuss linear birefringence in ribbons array of anisotropic 2D materials. We found, relative to the extended surface, that both phase retardation and amplitude attenuation of the scattered field can be altered more drastically in the array geometry. This includes scenarios where the array exhibits linear-circular polarization conversion in its reflected mode, while concurrently, acts as a polarization rotator in the transmitted mode. The wide tunability of the array's response, with simultaneous control of angular detuning and anisotropic Drude weights, renders anisotropic ribbons array as viable platforms to be used in dynamic metadevices.      


\emph{Acknowledgment} -- The authors acknowledge helpful discussions with Sang-Hyun Oh and In-Ho Lee. K. K. was supported primarily by the National Science Foundation through the University of Minnesota MRSEC under Award Number DMR-1420013. L. M. M. acknowledges the Spanish Ministry of Economy and Competitiveness under project MAT2014-53432-C5-1-R.


\bibliography{./BPLB}

\end{document}